\documentclass
[superscriptaddress,secnumarabic,amssymb,amsmath,nobibnotes,aps,prd,showkeys,showpacs,nofootinbib]{revtex4}%
\usepackage{graphicx}
\usepackage{epsf}
\usepackage{bm}
\usepackage{amsmath}
\usepackage{amsfonts}
\usepackage{amssymb}
\usepackage{epstopdf}%
\providecommand{\U}[1]{\protect\rule{.1in}{.1in}}
\newcommand{\be}{\begin{equation}}
\newcommand{\ee}{\end{equation}}

\newcommand{\mincir}{\raise
-3.truept\hbox{\rlap{\hbox{$\sim$}}\raise4.truept\hbox{$<$}\ }}
\newcommand{\magcir}{\raise
-3.truept\hbox{\rlap{\hbox{$\sim$}}\raise4.truept\hbox{$>$}\ }}

\begin{document}
\title{Noether symmetries and duality transformations in cosmology}
\author{Andronikos Paliathanasis}
\email{anpaliat@phys.uoa.gr}
\affiliation{Instituto de Ciencias F\'{\i}sicas y Matem\'{a}ticas, Universidad Austral de
Chile, Valdivia, Chile}
\author{Salvatore Capozziello}
\email{capozziello@na.infn.it}
\affiliation{Dipartimento di Fisica, Universita' di Napoli Federico II, Complesso
Universitario di Monte S. Angelo, Via Cinthia, 9, I-80126 Naples, Italy.}
\affiliation{Istituto Nazionale di Fisica Nucleare (INFN) Sez. di Napoli, Complesso
Universitario di Monte S. Angelo, Via Cinthia, 9, I-80126 Naples, Italy.}
\affiliation{Gran Sasso Science Institue (INFN), Viale F. Crispi 7, I-67100, L' Aquila, Italy.}

\begin{abstract}
We discuss the relation between Noether (point) symmetries and discrete
symmetries for a class of minisuperspace cosmological models. We show that
when a Noether symmetry exists for the gravitational Lagrangian then there
exists a coordinate system in which a reversal symmetry exists. Moreover as
far as concerns the scale-factor duality symmetry of the dilaton field, we
show that it is related to the existence of a Noether symmetry for the field
equations, and the reversal symmetry in the normal coordinates of the symmetry
vector becomes scale-factor duality symmetry in the original coordinates. In
particular the same point symmetry as also the same reversal symmetry exists
for the Brans-Dicke- scalar field with linear potential while now the discrete
symmetry in the original coordinates of the system depends on the Brans-Dicke
parameter and it is a scale-factor duality when $\omega_{BD}=1$. Furthermore,
in the context of the O'Hanlon theory for $f\left(  R\right)  $-gravity, it is
possible to show how a duality transformation in the minisuperspace can be
used to relate different gravitational models.

\end{abstract}
\date{\today}
\maketitle

\section{Introduction}

\label{intro}

The necessity to explain the new phenomena, which have been discovered from
the analysis of the recent cosmological data
\cite{Teg,Kowal,Komatsu,Suzuki,planck,planck2015}, has led to gravitational
theories which extend Einstein's General Relativity (GR). The modification%
$\backslash$%
extension of the Einstein-Hilbert action which keeps the linearity of the
gravitational action is the introduction of the cosmological constant,
$\Lambda$, which is the simplest candidate to explain the late-time
acceleration of the universe.

Inspired from the field theory, new scalar fields, minimally or nonminimally
coupled to gravity, either higher-order curvature invariants have been
introduced into the gravitational action, for instance see
\cite{faraonibook,CapR,odin1,odin2,Oliva,Vernov,Vernov1,PhysRep}. The main issue of modified%
$\backslash$%
extended theories of gravity is that the new components in the gravitational
field equations, which follow from the new terms in the gravitational action,
change the dynamics of the field equations such as the solutions of the latter
to describe the observable phenomena. The extra terms arising from the
modification of gravitational action can be seen as the components of a sort
of effective geometric dark energy momentum tensor \cite{capQ}. However, such
terms can be related to properties coming from fundamental physics.

In particular, an important property of the two-dimensional conformal field
theory and, consequently of the string theory, is the duality symmetry which
has important consequences in cosmology (for reviews see
\cite{Alvarez,review1}). As duality symmetry is characterized by the
invariance of the action integral, therefore the corresponding Euler-Lagrange
equations, under this transformation, remain the same. Usually when the
"radius" $\mathcal{R}$~of the geometry changes such as $\bar{\mathcal{R}%
}\rightarrow\mathcal{R}^{-1}$. Duality transformation is a discrete
transformation and an isometry should exist for the underlying manifold
\cite{Buscher1,Buscher2}. However isometries form a subalgebra on the
Homothetic algebra of a manifold, while the latter algebra is related to the
group of local invariant transformations which transforms the action integral
in such a way that the Euler-Lagrange equations are invariant. These are the
so called Noether (point) symmetries \cite{mtgrg}.

Furthermore, the dilaton scalar field model in a
Friedmann-Lema\^{\i}tre-Robertson-Walker spacetime (FLRW) admits a
scale-factor duality symmetry \cite{callan}, that is, when the scale-factor
changes such as $a\rightarrow$ $a^{-1}$, and the dilaton field is shifted with
a given formula, the gravitational action remains invariant. As we will see
below, in the minisuperspace approach, the dilaton action admits a Noether
conservation law, while the Noether symmetry and the conservation law are also
invariant under duality transformations. However Noether symmetries provide
point transformations where the field equations are invariant, despite of the
fact that the duality transformations are discrete transformations.

Mathematically the two identities are different, however as we will see, for a
class of cosmological models, the existence of a Noether conservation law
indicates the existence of a discrete transformation, i.e. of a discrete
symmetry,\ (not necessary that of scale-factor duality transformation).

This work focuses on the relation between discrete transformations following
from the existence of Noether symmetries in cosmological Lagrangians. As a
byproduct, we will show that duality transformations are a particular case of
these discrete transformations.

The plan of the paper is the following. In Section \ref{duality}, we briefly
discuss the duality transformation of the sigma model and the scale-factor
duality of the dilaton field. Using previous results for Noether symmetries in
minisuperspace, in Section \ref{noether}, we discuss the existence of Noether
conservation laws, i.e. local transformations, with the existence of discrete
transformations in the normal coordinates of the symmetry vector. In
particular, we show that the existence of the Homothetic vector field, not
necessary an isometry vector in the minisuperspace, generates at least a
Noether symmetry for the corresponding Lagrangian. It is equivalent to the
existence of a discrete transformation which leaves invariant the dynamical
system. This approach is used in Section \ref{application1}, where we show
that the Brans-Dicke scalar field, with linear potential, admits scale-factor
duality symmetry for the\ Brans-Dicke parameter $\omega_{BD}=1$. The
Brans-Dicke field corresponds to the dilaton field also for models with
arbitrary parameter $\omega_{BD}$ admitting the same number of Noether
symmetries. However the transformation from the original coordinates to the
normal coordinates depends on $\omega_{BD}$, and the reverse symmetry, in the
normal coordinates, becomes scale-factor duality in the original coordinates
if and only if $\omega_{BD}=1$. Moreover, another kind of discrete
transformation exists for other values of the Brans-Dicke parameter. Finally,
we consider the O'Hanlon theory, which is equivalent to the $f\left(
R\right)  $-gravity. We find a family of potentials that can be related each
other under a discrete transformation which is a duality transformation in the
minisuperspace. In Section \ref{conclusion}, we draw conclusions.

\section{String Duality and Scale factor Duality}

\label{duality}

Following Buscher \cite{Buscher1,Buscher2} we consider the general dualizable
bosonic non-linear sigma model on a $D-$dimensional manifold $\mathcal{M~}$of
Lorentzian signature with a dilaton field $\phi~$coupled to the curvature
scalar of the two-dimensional metric tensor $\gamma_{\mu\nu}$. The action
integral of the latter model can be written in the following form%
\begin{equation}
S=\frac{1}{4\pi a^{\prime}}\int d^{2}\xi\left(  \sqrt{\gamma}\gamma^{\mu\nu
}g_{ab}\partial_{\mu}x^{\alpha}\partial_{\nu}x^{b}+\varepsilon^{\mu\nu}%
h_{ab}\partial_{\mu}x^{\alpha}\partial_{\nu}x^{b}+a^{\prime}\sqrt{\gamma
}R^{\left(  2\right)  }\phi\left(  x^{\gamma}\right)  \right)  , \label{d.01}%
\end{equation}
where $g_{ab}$ is the metric tensor of the target space, $h_{ab}$ is the
torsion, and $a^{\prime}$ is the inverse of the string tension. We assume that
the manifold $M$, admits an isometry vector field, specifically a translation
symmetry, that is the $\mathcal{M}$ manifold can seen as $1+\left(
D-1\right)  $ space.

Hence, the dualized model has the following action integral%
\begin{equation}
\bar{S}=\frac{1}{4\pi a^{\prime}}\int d^{2}\xi\left(  \sqrt{\gamma}\gamma
^{\mu\nu}\bar{g}_{ab}\partial_{\mu}\bar{x}^{\alpha}\partial_{\nu}\bar{x}%
^{b}+\varepsilon^{\mu\nu}\bar{h}_{ab}\partial_{\mu}\bar{x}^{\alpha}%
\partial_{\nu}\bar{x}^{b}+a^{\prime}\sqrt{\gamma}R^{\left(  2\right)  }%
\phi\left(  x^{\gamma}\right)  \right)  , \label{d.02}%
\end{equation}
where the new metric tensor $\bar{g}_{ab}$, and the new torsion $\bar{h}_{ab}$
are related with that of (\ref{d.01}) as follows%
\begin{equation}
\bar{g}_{00}=\left(  g_{00}\right)  ^{-1}~,~\bar{g}_{0i}=h_{0i}\left(
g_{00}\right)  ^{-1}~,~\bar{g}_{ij}=g_{ij}-\left(  g_{0i}g_{0j}-h_{0i}%
h_{0j}\right)  \left(  g_{00}\right)  ^{-1}, \label{d.03}%
\end{equation}%
\begin{equation}
\bar{h}_{0i}=-h_{i0}=g_{0i}\left(  g_{00}\right)  ^{-1}~,~\bar{h}_{ij}%
=h_{ij}+\left(  g_{0i}h_{0j}-h_{0i}g_{0j}\right)  \left(  g_{00}\right)
^{-1}, \label{d.04}%
\end{equation}
in which~$i=1,2,3...D.$

The two action integrals (\ref{d.01}) and (\ref{d.02}), through the
variational principle, are equivalent at classical level. However, in order to
be equivalent at the quantum level, the action integral (\ref{d.01}) has to be
conformally invariant. Conformal invariance of (\ref{d.01}), at the one-loop
level, gives that the dilaton field $\phi$ has to satisfy the following
conditions \cite{callan}%
\begin{equation}
\frac{1}{a^{\prime}}\frac{D-26}{48\pi^{2}}+\frac{1}{16\pi^{2}}\left(  4\left(
\nabla\phi\right)  ^{2}-4\nabla^{2}\phi-R+\frac{1}{12}H^{2}\right)  =0,
\label{d.06}%
\end{equation}%
\begin{equation}
R_{ab}-\frac{1}{4}H_{a}^{cd}H_{bcd}+2\nabla_{a}\nabla_{b}\phi=0, \label{d.07}%
\end{equation}%
\begin{equation}
\nabla_{c}H_{ab}^{c}-2\left(  \nabla_{c}\phi\right)  H_{ab}^{c}=0,
\label{d.08}%
\end{equation}
where $R_{ab}$ is the Ricci tensor related to the metric tensor $g_{ab},$ $R$
is the Ricci scalar, $H_{abc}=3\nabla_{\lbrack a}h_{bc]}$ is the antisymmetric
tensor strength, and $H^{2}=H_{abc}H^{abc}$. \ Therefore the dualized model
(\ref{d.02}) is conformally invariant if dilaton field shifts as
\begin{equation}
\bar{\phi}=\phi-\frac{1}{2}\ln\left(  g_{00}\right)  . \label{d.09}%
\end{equation}
The set of discrete transformations which relates the two models (\ref{d.01}),
(\ref{d.02}) form the duality transformations of the theory. Furthermore, the
transformation is an element of the discrete symmetry group $O\left(
d,d\right)  $~\cite{Shapere}.

\subsection{Scale-factor duality}

The equations that the dilaton field $\phi$ has to satisfy, i.e. Eqs
(\ref{d.06})-(\ref{d.08}), follow from the variation principle of the action
integral\footnote{In the following we consider signature $\left(
-++...\right)  $.}%
\begin{equation}
S_{dilaton}=s_{0}\int d^{D}x\sqrt{\left\vert g\right\vert }e^{-2\phi}\left(
R-4\nabla_{a}\phi\nabla^{a}\phi-\frac{1}{12}H^{2}-\Lambda\right)  .
\label{d.10}%
\end{equation}
where the cosmological constant is $\Lambda=\frac{1}{a^{\prime}}\frac
{D-26}{3\pi^{2}}$, and $s_{0}$ is a rescaling constant. \ Since we are
interested in cosmology, we can consider the antisymmetric parts vanishing,
i.e. $H^{2}=0$. Therefore the action integral (\ref{d.10}) takes the simpler
form,
\begin{equation}
S_{dilaton}=s_{0}\int d^{D}x\sqrt{\left\vert g\right\vert }e^{-2\phi}\left(
R-4\nabla_{a}\phi\nabla^{a}\phi-\Lambda\right)  , \label{d.11}%
\end{equation}
which gives rise to an equivalent scalar-tensor model \cite{faraonibook}%
\begin{equation}
S_{ST}=\int d^{D}x\sqrt{\left\vert g\right\vert }\left[  f\left(  \phi\right)
R-\omega\left(  \phi\right)  \nabla_{a}\phi\nabla^{a}\phi-V\left(
\phi\right)  \right]  \label{d.12}%
\end{equation}
with $f\left(  \phi\right)  =4^{-1}\omega\left(  \phi\right)  =e^{-2\phi}$,
and $V\left(  \phi\right)  =\Lambda f\left(  \phi\right)  $.

In the context of a $D$-dimensional spatially flat FLRW with line element%
\begin{equation}
ds^{2}=-dt^{2}+a^{2}\left(  t\right)  \delta_{\alpha\beta}dx^{\alpha}%
dx^{\beta}, \label{d.13}%
\end{equation}
it has been shown that there exist scale-factor duality. Specifically the
action integral (\ref{d.10}) and the field equations (\ref{d.06})-(\ref{d.08})
are invariant under the discrete transformation\footnote{The scale-factor
duality is a symmetry for (\ref{d.10}) with or without the torsion term
\cite{Gasp}.} \cite{Gasp}
\begin{equation}
\bar{a}=a^{-1}~,~\bar{\phi}=\phi-\left(  D-1\right)  \ln a, \label{d.14}%
\end{equation}
which is of the form of Buscher's duality transformation, that is, the dilaton
field admits a duality symmetry.

Consider now that the dimension of the FLRW spacetime is four, $D=4$, that is,
the Ricci scalar is given as follows,%
\begin{equation}
R=6\left[  \frac{\ddot{a}}{a}+\left(  \frac{\dot{a}}{a}\right)  ^{2}\right]
.\label{d.15}%
\end{equation}
By replacing (\ref{d.15}) in the action integral (\ref{d.11}) the point
Lagrangian is
\begin{equation}
L\left(  a,\dot{a},\phi,\dot{\phi}\right)  =e^{-2\phi}\left(  6a\dot{a}%
^{2}-12a^{2}\dot{a}\dot{\phi}+4a^{3}\dot{\phi}^{2}-\Lambda a^{3}\right)
,\label{d.16}%
\end{equation}
where the field Eqs (\ref{d.06})-(\ref{d.08}) can be written as
\begin{equation}
e^{-2\phi}\left(  6a\dot{a}-12a^{2}\dot{a}\dot{\phi}+4a^{3}\dot{\phi}%
^{2}-\Lambda\right)  =0,\label{d.17}%
\end{equation}%
\begin{equation}
\ddot{a}+\frac{2}{a}\dot{a}^{2}-2\dot{a}\dot{\phi}=0,\label{d.18}%
\end{equation}%
\begin{equation}
\ddot{\phi}+\frac{3}{2a^{2}}\dot{a}^{2}-\dot{\phi}^{2}-\frac{\Lambda}%
{4}=0.\label{d.19}%
\end{equation}
However it is not difficult to observe that the Brans-Dicke scalar field,
\begin{equation}
S_{BD}=\int d^{D}x\sqrt{\left\vert g\right\vert }\left(  \psi R-\frac
{\omega_{BD}}{\psi}\nabla_{a}\psi\nabla^{a}\psi-V_{0}\psi\right)  \label{d.20}%
\end{equation}
with $\omega_{BD}=1,~$admits the scale-factor duality symmetry, while now the
duality transformation is
\begin{equation}
\bar{a}=a^{-1}~,~~\bar{\psi}=\psi a^{6}.\label{d.21}%
\end{equation}
The reason of this result is that the two action integrals (\ref{d.11}) and
(\ref{d.20}) are related under the change of variables/fields, $\phi=-\frac
{1}{2}\ln\psi$, when the Brans-Dick parameter is $\omega_{BD}=1$. Of course
that property holds for all the scalar-tensor theories (\ref{d.12}) where a
new filed $\Psi$ can be defined, such as the action is (\ref{d.11}). In that
case the scale-factor duality symmetry exists, where the scalar field has to
be shifted according to the formula connecting the field $\Psi$, and the
dilaton field $\phi$. Specifically all the scalar tensor models (\ref{d.12})
with functions%
\begin{equation}
f\left(  \phi\right)  =4^{-1}e^{-2\Omega\left(  \phi\right)  }~,~\omega\left(
\phi\right)  =\left(  \Omega_{,\phi}\right)  ^{2}e^{-2\Omega\left(
\phi\right)  }~,~V\left(  \phi\right)  =\Lambda f\left(  \phi\right)
\end{equation}
are invariant under a scale factor duality transformation (see also
\cite{dualityCap2}).

Furthermore for the Brans-Dicke scalar field (\ref{d.20}) it has been found
that the field equations admits at least one Noether conservation law
\cite{wdwBD}, with or without a dust fluid term. Because the conservation
laws, and the corresponding symmetries, are independent of the
\textquotedblleft coordinate\textquotedblright\ systems, it means that the
Lagrangian (\ref{d.16}) of the dilaton field admits at least a Noether
symmetry. \ The corresponding Noether point symmetry of Lagrangian
(\ref{d.16}) is given by the following vector field
\begin{equation}
X_{N}=a\partial_{a}+\frac{3}{2}\partial_{\phi}, \label{d.22a}%
\end{equation}
where the corresponding Noether conservation law is%
\begin{equation}
I=6a^{2}e^{-2\phi}\dot{a}, \label{d.22}%
\end{equation}
which is also invariant under the scale-factor duality transformation
(\ref{d.14}). This is not surprising since the form of the Lagrangian is the
same, before or after the discrete transformation, but it is worthy to mention
that either the sign of the conservation law does not change.
Furthermore as we will see in Sec. (\ref{application1}), \ Lagrangians
(\ref{d.16}) and (\ref{d.20}) are maximally symmetric which means that they
admit eight Noether point symmetries as many as the free particle case in a
two dimensional spacetime, or the two dimensional \textquotedblleft
oscillator\textquotedblright. However since the potential of the Lagrangian is
not constant, as we will see in the following, this fact indicates that the
classical equivalent dynamical system, which is described by (\ref{d.16}) or
(\ref{d.20}), is the \textquotedblleft oscillator\textquotedblright\ in the
two dimensional flat space with Lorentzian signature\footnote{We
remark that the Noether symmetries are transformed as vector fields under
coordinate transformations, that is, if  $J$  is the Jacobian of the
coordinate transformation then, in the new coordinates, the symmetry vector
$X$, is given as $\bar{X}=J\left(  X\right)$.}. 

The application of Noether Symmetry Approach to gravitational theories is
discussed in the following section. We show that for a class of cosmological
models the existence of a Noether conservation law indicates the existence of
a discrete transformation, i.e. discrete symmetry.

\section{Noether symmetries and discrete transformations}

\label{noether}

Emmy Noether \cite{EmmyN} proved that, under a point transformation with
generator $X=\xi\left(  t,x^{k}\right)  \partial_{t}+\eta^{i}\left(
t,x^{k}\right)  \partial_{i}$, the action integral of a Lagrangian function
$L\left(  t,x^{k},\dot{x}^{k}\right)  ~$transforms in a way such the following
condition holds%
\begin{equation}
X^{\left[  1\right]  }L+L\dot{\xi}=\dot{f} \label{d.23}%
\end{equation}
then the Euler-Lagrange equations are invariants\footnote{$X^{\left[
1\right]  }=X+\left(  \dot{\eta}^{i}-\dot{x}^{i}\dot{\xi}\right)
\partial_{\dot{x}^{i}}$ indicates the first prologation/extension of $X$ in
the space of variables $\left\{  t,x^{i},\dot{x}^{i}\right\}  ~$\cite{Ibra},
and dot means derivative with respect to \textquotedblleft$t$%
\textquotedblright.}, and $X$ is called Noether (point) symmetry\footnote{From
(\ref{d.23}) it is clear that it is not necessary the Lagrangian to be
invariant under the point transformation. However a function $\dot{f}$ can be
introduced, where the latter is eliminated by the variational principle.}. The
second part of Noether's theorem states that if condition (\ref{d.23}) holds
for a given vector field $X$, then the following quantity is a conservation
law for the Euler-Lagrange equations,%
\begin{equation}
I=\xi\left(  \dot{x}^{k}\frac{\partial L}{\partial\dot{x}^{k}}-L\right)
-\eta^{k}\frac{\partial L}{\partial\dot{x}^{k}}+f. \label{d.24}%
\end{equation}

We remark that Noether's theorem extends for higher-order systems or for
non-pointlike symmetries, however the above definition is the one that we are
interested in this work.

Moreover, from the point symmetry $X$, the point transformation can be derived
from the solution of the following system%
\begin{equation}
t^{\prime}=t+\varepsilon\xi\left(  t,x^{k}\right)  ~\ ~,~~x^{\prime i}%
=x^{i}+\varepsilon\eta^{i}\left(  t,x^{k}\right)  , \label{d.25}%
\end{equation}
where $\varepsilon$ is an infinitesimal parameter of the point transformation.

We are interested on scalar field, mutiple-scalar field, scalar-tensor
theories, and higher-order gravities such as,~$f\left(  R\right)  ,~f\left(
R,\square R,...\right)  ,$ where the Lagrangian of the field equations can be
written as follows, (either with the use of Lagrange multiplier)%
\begin{equation}
L\left(  x^{k},\dot{x}^{k}\right)  =\frac{1}{2}\gamma_{ij}\dot{x}^{i}\dot
{x}^{j}-U\left(  x^{k}\right)  , \label{d.26}%
\end{equation}
where $\gamma_{ij}$, indicates the minisuperspace of the cosmological model,
$i=$~$\#_{spacetime}^{variables}$ $+\#_{fields}^{variables}$, and $U\left(
x^{k}\right)  $ is the effective potential of the field equations, including
the curvature components, which corresponds to the spacetime and the dynamic
quantities of the new fields and matter terms when they are considered in the
discussion. Furthermore, the gravitational field equations, which corresponds
to the model with Lagrangian (\ref{d.26}), are the set of the Hamiltonian
function and the Euler-Lagrange equations,~$E_{L}\left(  L\right)  =0$.
Finally, we have to point out that $L\left(  x^{k},\dot{x}^{k}\right)  $ is
autonomous. The Noether symmetry condition (\ref{d.23}) for Lagrangians of the
form (\ref{d.26}) have been discussed in \cite{mtgrg}, and it has been found
that the component $\eta^{i}$ of the symmetry vector $X$ is an element of the
Homothetic algebra of $\gamma_{ij}$, while $\xi$ is analogue to the Homothetic
factor $\psi$ of the homothetic vector field. As we have mentioned, Lagrangian
(\ref{d.26}) is autonomous, which means that the symmetry vector $\partial
_{t}$ exists and, at the same time, the field equations are invariant under
time-reversal transformation,i.e.$~t\rightarrow-t~$\cite{Gasp}. Following
\cite{mtgrg} and for the completeness of our analysis we consider the two
cases $\dot{f}=0$,~and $\dot{f}\neq0$ for the symmetry condition (\ref{d.23}).

Let us consider $\dot{f}=0$ in (\ref{d.23}), then, for the Lagrangian
(\ref{d.26}), the Noether symmetry has the form, $X=2\psi t\partial_{t}%
+H^{i}\partial_{i},$ and the following condition holds%
\begin{equation}
\mathcal{L}_{H}U+2\psi U=0. \label{d.27}%
\end{equation}
where $H^{i}$ is the Homothetic vector of the tensor $\gamma_{ij}$, i.e.
$\mathcal{L}_{H}\gamma_{ij}=2\psi\gamma_{ij}$, and, when $\psi=0$, $H^{i}$ is
an isometry (or a Killing vector).When $H^{i}$ is an isometry then, from
(\ref{d.27}) and the Killing condition, we have that there exist a coordinate
system $\left\{  x^{i}\right\}  \rightarrow\left\{  y^{I},y^{A}\right\}  $,
such as the isometry becomes $\partial_{I}~$(these are called normal
coordinates), and the Lagrangian (\ref{d.26}) becomes, for a nonnull isometry
\cite{dualitydilaton},
\begin{equation}
L\left(  y^{k},\dot{y}^{k}\right)  =\frac{1}{2}\left(  \gamma_{II}\left(
y^{A}\right)  \left(  \dot{y}^{I}\right)  ^{2}+\gamma_{AB}\left(
y^{A}\right)  \dot{y}^{A}\dot{y}^{B}\right)  -U\left(  y^{A}\right)  ,
\label{d.28}%
\end{equation}
where $A\neq I$~and $i=I,A$. However when the isometry is a null vector,
i.e.~$H^{i}H_{i}=0$, in the normal coordinates, Lagrangian (\ref{d.26})
becomes \cite{julia}%
\begin{equation}
L\left(  u,\dot{u},v,\dot{v},y^{\beta},\dot{y}^{\beta}\right)  =\frac{1}%
{2}\left(  \gamma_{uu}\left(  y^{\beta},u\right)  du^{2}+2\gamma_{uv}\left(
y^{\beta},u\right)  dudv+\gamma_{\alpha\beta}\left(  y^{\beta},u\right)
\dot{y}^{\alpha}\dot{y}^{\beta}\right)  -U\left(  y^{\beta},u\right)  ,
\label{d.28b}%
\end{equation}
where the null isometry and the new coordinates are $\partial_{v}$ and
$\left\{  x^{i}\right\}  \rightarrow\left\{  u,v,y^{\beta}\right\}  .$

It is straightforward to see that, under the discrete \textquotedblleft
space\textquotedblright\ reversal transformation $y^{I}\rightarrow-y^{I}$,
Lagrangian (\ref{d.28}) is always invariant. On the other hand, this is not
always true for (\ref{d.28b}): it is invariant under the reversal
transformation $\left\{  u,v\right\}  \rightarrow\left\{  -u,-v\right\}  $
only for $\gamma_{ij}\left(  y^{\beta},-u\right)  =\gamma_{ij}\left(
y^{\beta},u\right)  $ and $U\left(  y^{\beta},-u\right)  =U\left(  y^{\beta
},u\right)  $.

Now as far as concerns the original coordinate system, the reversal symmetry
can be different and depends on the coordinate transformation. As we will see
in the followings, the scale-factor duality transformation of the dilaton
Lagrangian (\ref{d.16}) is a reversal symmetry in the normal coordinates of
the dynamical system.

On the other hand, when $\psi\neq0$, from (\ref{d.27}) and the Homothetic
equation, we have that a coordinate system $\left\{  x^{i}\right\}
\rightarrow\left\{  Y^{I},Y^{A}\right\}  $ exists where the Homothetic vector
becomes $Y^{I}\partial_{I}$, while the Lagrangian (\ref{d.26}) takes the
following form%
\begin{equation}
L\left(  Y^{k},\dot{Y}^{k}\right)  =\frac{1}{2}\left(  \gamma_{II}\left(
Y^{A}\right)  \left(  \dot{Y}^{I}\right)  ^{2}+\left(  Y^{I}\right)
^{2}\gamma_{AB}\left(  Y^{A}\right)  \dot{Y}^{A}\dot{Y}^{B}\right)
-\frac{\bar{U}\left(  Y^{A}\right)  }{\left(  Y^{I}\right)  ^{2}},
\label{d.29}%
\end{equation}
that is invariant under the reversal \textquotedblleft
coordinate\textquotedblright\ transformation $Y^{I}\rightarrow-Y^{I}$.

If $\dot{f}\neq0$ in the symmetry condition (\ref{d.23}), the generic Noether
symmetry vector for (\ref{d.26}) is $X=2\psi\int T\left(  t\right)
\partial_{t}+T\left(  t\right)  S^{,i}\partial_{i}$, when the following
condition
\begin{equation}
\mathcal{L}_{S}U+2\psi U+\sigma S=0, \label{d.30}%
\end{equation}
holds, where now $S^{,i}$ is a gradient Homothetic symmetry (or isometry) and
function $T\left(  t\right)  $ is given by the second-order differential
equation $T_{,tt}=\sigma T$.

In \cite{Ermakov}, the cases when $U\left(  x^{i}\right)  $ does not admit
linear terms of the gradient function of the Killing isometry vector $S_{KV}$
have been discussed. If $S^{,i}$, in our notation, is a gradient (non-null)
isometry vector field, in the normal coordinates $\left\{  y^{i}\right\}  ,$
Lagrangian (\ref{d.26}) can be written with oscillatory terms%
\begin{equation}
L\left(  y^{k},\dot{y}^{k}\right)  =\frac{1}{2}\left(  \left(  \dot{y}%
^{I}\right)  ^{2}+\gamma_{AB}\left(  y^{A}\right)  \dot{y}^{A}\dot{y}%
^{B}\right)  -\mu\left(  y^{I}\right)  ^{2}-\bar{U}\left(  y^{A}\right)  ,
\label{d.31}%
\end{equation}
Finally, if $S^{,i}$ is a gradient Homothetic symmetry in the normal
coordinates $\left\{  Y^{i}\right\}  $, Lagrangian (\ref{d.26}) describes the
generalized Riemannian Ermakov-Pinney system%
\begin{equation}
L\left(  Y^{k},\dot{Y}^{k}\right)  =\frac{1}{2}\left(  \left(  \dot{Y}%
^{I}\right)  ^{2}+\left(  Y^{I}\right)  ^{2}\gamma_{AB}\left(  Y^{A}\right)
\dot{Y}^{A}\dot{Y}^{B}\right)  -\mu\left(  Y^{I}\right)  ^{2}-\frac{\bar
{U}\left(  Y^{A}\right)  }{\left(  Y^{I}\right)  ^{2}}. \label{d.32}%
\end{equation}
Therefore Lagrangians (\ref{d.31}), (\ref{d.32}) are invariant under the
discrete transformations $y^{I}\rightarrow-y^{I}$ and $Y^{I}\rightarrow-Y^{I}%
$. Furthermore, we observe that if $\mu=0$, and $\gamma_{II}=1$, then
Lagrangians (\ref{d.31}), (\ref{d.32}) are equivalent to Lagrangians
(\ref{d.28}) and (\ref{d.29}) respectively.

Now, if the minisuperspace $\gamma_{ij}$ admits a gradient isometry and the
potential $U\left(  x^{i}\right)  ~$admits linear terms of $S_{KV}$, then, in
the normal coordinates $\left\{  y^{i}\right\}  $, it is $S_{KV}=y^{I}$, and,
in the normal coordinates, the effective potential is $U\left(  y^{i}\right)
=\sigma y^{I}+\bar{U}\left(  y^{A}\right)  $, $\sigma\neq0$. In this case a
Noether conservation laws exists and, however, the reversal symmetry does not exist.

Of course, the above analysis holds for a single symmetry vector acting on the
Lagrangian. If more than one symmetries exist, then new discrete
transformations can be found. However it is worth stressing that, while in one
coordinate system the discrete transformation can be seen as a reversal
transformation, in another coordinate system, it can be a different transformation.

In the case where the Noether symmetry is generated by a null isometry, i.e.
the Lagrangian has the form of (\ref{d.28b}) for $\gamma_{uu}=0$, then, under
the transformation $\left\{  u\rightarrow v,v\rightarrow u\right\}  $, the
dynamical system is not necessary invariant. However the two dynamical systems
are related by a simple change of variables. When the elements of the
metric~$\gamma_{ij},~$\ i.e. $\gamma_{uv}\left(  y^{\nu},u\right)  $,
$\gamma_{ij}\left(  y^{\nu},u\right)  ,~$ and the potential $\bar{U}\left(
y^{\beta},u\right)  $ are not dependent on the variable $u$, then the
dynamical system is invariant. If that is true, then it is easy to show that
the dynamical system admits more than one Noether conservation laws.

In the following section, we apply the above results to some cosmological
models where Noether point symmetries can be identified.

\section{Duality transformations from Noether symmetries}

\label{application1}

Let us consider a simple well-known system in $\mathcal{M}^{2}$ space which
follows from the Lagrangian%
\begin{equation}
L\left(  x,\dot{x},y,\dot{y}\right)  =\frac{\Omega}{2}\left(  \dot{x}^{2}%
-\dot{y}^{2}\right)  -\frac{\mu}{2}\left(  x^{2}-y^{2}\right)  , \label{d.33}%
\end{equation}
where $\Omega,\mu$ are constants. Lagrangian (\ref{d.33}) describes the
two-dimensional \textquotedblleft oscillator\textquotedblright\ in a flat
space with Lorentzian signature. It is straightforward to see that
(\ref{d.33}) admits eight Noether point symmetries (the Noether symmetries can
be found for instance in \cite{Sen,Damianou}). Also the dynamical system in
which Lagrangian (\ref{d.33}) describes is invariant under the following
discrete transformations $\left\{  x\rightarrow-x\right\}  $,~or~$\left\{
y\rightarrow-y\right\}  $,~and the complex one $\left\{  x\rightarrow
iy~,~y\rightarrow ix\right\}  $, while, under the transformation $\left\{
x\rightarrow y,~y\rightarrow x\right\}  $, the new Lagrangian $\bar{L}$, is
$\bar{L}=-L$. However while the field equations have same the sign, the
Hamiltonian constant changes: if the latter is zero then everything is
invariant. As we can see Lagrangian (\ref{d.33}) is of the form of Lagrangian
(\ref{d.31}). Hence these discrete symmetries for Lagrangian
(\ref{d.33}) follows directly from the existence of the Noether symmetry
vectors as we discussed in Sec. \ref{noether}. 

Specifically the transformations $\left\{  x\rightarrow-x\right\}
$,~$\left\{  y\rightarrow-y\right\}  $ follow from the translation
symmetries while the remaining two transformations follow from the rotation
symmetry and the homothetic symmetry of  $M^{2}$ space. As we will
show below, these discrete symmetries while are reversal symmetries in the
Cartesian coordinates, in the coordinates where the dilaton field is defined,
take the form of the scale factor duality symmetry.

Consider now the coordinate transformation,%
\begin{equation}
x=\frac{\sqrt{2}}{2}\left(  u+v\right)  ~,~y=\frac{\sqrt{2}}{2}\left(
u-v\right)  , \label{d.34}%
\end{equation}
with inverse
\begin{equation}
u=\frac{\sqrt{2}}{2}\left(  x+y\right)  ~,~v=\frac{\sqrt{2}}{2}\left(
x-y\right)  . \label{d.35}%
\end{equation}

Hence, the discrete transformation $\left\{  x\rightarrow-x,y\rightarrow
-y\right\}  $ \ in the new coordinates system corresponds to $\left\{
u\rightarrow-u~,~v\rightarrow-v\right\}  $, while the transformation $\left\{
x\rightarrow-x\right\}  $ corresponds to $\left\{  u\rightarrow
-v~,~v\rightarrow-u\right\}  $, and $\left\{  y\rightarrow-y\right\}  $
corresponds to the discrete transformation $\left\{  u\rightarrow
v~,~v\rightarrow u\right\}  $. \ All these transformations are related with
the admitted translation group of the flat space.

In the new coordinate system $\left\{  u,v\right\}  $, Lagrangian (\ref{d.33})
becomes%
\begin{equation}
L\left(  u,\dot{u},v,\dot{v}\right)  =\Omega\left(  \dot{u}\dot{v}\right)
-\mu\left(  uv\right)  . \label{d.36}%
\end{equation}
We assume the second coordinate transformation%
\begin{equation}
a=u^{p_{+}}v^{p_{-}}~,~\phi=\ln\left(  u^{q_{+}}v^{q_{-}}\right)  ,
\label{d.37}%
\end{equation}
in which the constants $p_{\pm}$,~$q_{\pm}$ are%
\begin{equation}
p_{\pm}=\frac{\left(  \omega-4\right)  \sqrt{6\left(  6-\omega\right)  }%
\pm4\left(  \omega-6\right)  }{\sqrt{6\left(  6-\omega\right)  }\left(
3\omega-16\right)  }, \label{d.38}%
\end{equation}%
\begin{equation}
q_{\pm}=2\frac{\sqrt{6\left(  6-\omega\right)  }\pm3\left(  \omega-6\right)
}{\sqrt{6\left(  6-\omega\right)  }\left(  3\omega-16\right)  }. \label{d.39}%
\end{equation}
where
\begin{equation}
\Omega=8\frac{\left(  \omega-6\right)  }{3\omega-16}. \label{d.39b}%
\end{equation}
Then under the second transformation, Lagrangian (\ref{d.36}) becomes
\begin{equation}
L\left(  a,\dot{a},\phi,\dot{\phi}\right)  =e^{-2\phi}\left(  6a\dot{a}%
^{2}-12a^{2}\dot{a}\dot{\phi}+\omega a^{3}\dot{\phi}^{2}-\Lambda a^{3}\right)
, \label{d.40}%
\end{equation}
where the constant $\Lambda=\mu$. The latter Lagrangian has the form of
(\ref{d.16}) with a difference in the coefficient of $\dot{\phi}^{2}$.
Furthermore, it is easy to see that, under the transformation $\phi=-\frac
{1}{2}\ln\left(  \psi\right)  $, Lagrangian (\ref{d.40}) becomes of
Brans-Dicke form as (\ref{d.20}) where $\omega=4\omega_{BD}$. \ 

From transformations (\ref{d.37}), we see that the discrete transformation
$\left\{  y\rightarrow-y\right\}  $, in the Cartesian coordinates, or
$\left\{  u\rightarrow v~,~v\rightarrow u\right\}  $ in the coordinates
$\left\{  u,v\right\}  $, is that of the scale-factor duality%
\begin{equation}
a\rightarrow a^{-1}%
\end{equation}
if and only if $p_{+}=-p_{-}$. Using (\ref{d.38}), we find the unique solution
$\omega=4$, or $\omega_{BD}=1$, where (\ref{d.39}) corresponds to the
Lagrangian of the dilaton field (\ref{d.16}). Therefore we conclude that the
scale-factor duality transformation is related to the existence of Noether
symmetries of the Lagrangian and the field equations. The reason why only the
constant $\omega_{BD}=1$ is admitted for the a scale-factor duality follows
from the property of the coordinate transformation (\ref{d.37}).

In general, from (\ref{d.37}), we can achieve the inverse transformation
\begin{equation}
u=a^{Q_{-}}\exp\left(  -P_{-}\phi\right)  ,~v=a^{-Q_{+}}\exp\left(  P_{+}%
\phi\right)  , \label{d.41}%
\end{equation}
where the new constants are
\begin{equation}
P_{\pm}=\frac{p_{\pm}}{p_{+}q_{-}-p_{-}q_{+}}~,~Q_{\pm}=\frac{q_{\pm}}%
{p_{+}q_{-}-p_{-}q_{+}}~, \label{d.42}%
\end{equation}
that is, it is possible to construct discrete transformations where the
scalar-tensor model (\ref{d.40}) is invariant under Noether symmetries.

Hence, by using (\ref{d.37}) and (\ref{d.41}), we have that the discrete
transformation $\left\{  u\rightarrow v~,~v\rightarrow u\right\}  $ in the
non-diagonal coordinates $\left\{  u,v\right\}  $, the coordinates $\left\{
a,\phi\right\}  $ become%
\begin{equation}
a\rightarrow a^{\left(  p_{-}Q_{-}-p_{+}Q_{+}\right)  }\exp\left(  \left(
p_{+}P_{+}-p_{-}P_{-}\right)  \phi\right)  ,~ \label{d.43}%
\end{equation}%
\begin{equation}
\exp\left(  \phi\right)  \rightarrow a^{\left(  q_{-}Q_{-}-q_{+}Q_{+}\right)
}\exp\left(  \left(  q_{+}P_{+}-q_{-}P_{-}\right)  \phi\right)  , \label{d.44}%
\end{equation}
and Lagrangian (\ref{d.40}) is invariant. Transformation (\ref{d.43}),
(\ref{d.44}) is a discrete symmetry for (\ref{d.40}). Furthermore, as far as
concerns the Brans-Dicke field (\ref{d.20}), the discrete transformation which
does not change the field equations is%
\begin{equation}
a\rightarrow a^{\left(  p_{-}Q_{-}-p_{+}Q_{+}\right)  }\psi^{-\frac{1}%
{2}\left(  p_{+}P_{+}-p_{-}P_{-}\right)  }, \label{d.45}%
\end{equation}%
\begin{equation}
\psi\rightarrow a^{-2\left(  q_{-}Q_{-}-q_{+}Q_{+}\right)  }\psi^{\left(
q_{+}P_{+}-q_{-}P_{-}\right)  }. \label{d.46}%
\end{equation}
In order to clarify differences between discrete and local transformations,
let us consider the Noether symmetry vector (\ref{d.22a}) for the Brans-Dicke
scalar field (\ref{d.20}). Then, from (\ref{d.25}), we find the point
transformation
\begin{equation}
t\rightarrow t~,~a\rightarrow e^{\varepsilon}a~,~\phi\rightarrow\phi+\frac
{3}{2}\varepsilon, \label{d.46a}%
\end{equation}
which means that a rescaling of the scale factor keeps invariant, not only the
Lagrangian, but also the action integral if the field $\phi$ shifts properly.
Of course, for the remaining Noether symmetries of the model, new point
transformations can be constructed. However, while from a Noether
symmetry we can always construct a continuous transformation from the system
(\ref{d.25}) in order to find the discrete transformation, which leaves the
field equations invariant, we have to write the latter in the normal
coordinates of the symmetry vector, where there the discrete symmetry is a
reversal symmetry.

Furthermore, in order the reversal symmetry to provide a discrete
symmetry in the original coordinates of the dynamical system, the reversal
transformation should survive. To demonstrate this fact, let us consider the
\textquotedblleft polar\textquotedblright\ coordinates of (\ref{d.33}) and let
$\Omega=1,$ it is %
\begin{equation}
L\left(  r,\dot{r},\theta,\dot{\theta}\right)  =\frac{1}{2}\left(  \dot{r}%
^{2}-r^{2}\dot{\theta}^{2}\right)  -\frac{\mu}{2}r^{2}, \label{d.46b}%
\end{equation}
from where we can see that the latter Lagrangian is invariant under
the discrete transformations $\left\{  r\rightarrow-r\right\}  $\textbf{,
}$\left\{  \theta\rightarrow-\theta\right\}  $, while $\theta
\rightarrow-\theta$, is a transformation which follows from the
$\left\{  x\rightarrow-x\right\}  $\textbf{ or }$\left\{  y\rightarrow
-y\right\}  $. If we assume the second coordinate transformation
$r^{2}=\sqrt{\frac{8}{3}}a^{3}$,~$\theta=\sqrt{\frac{3}{8}}\Phi$,
then (\ref{d.46b}) becomes%
\begin{equation}
L\left(  a,\dot{a},\Phi,\dot{\Phi}\right)  =3a\dot{a}^{2}-\frac{1}{2}a^{3}%
\dot{\Phi}^{2}+a^{3}\Lambda\label{d.47c}%
\end{equation}
where $\mu=-2\sqrt{\frac{3}{8}}\Lambda$. From this fact, we
can see that, from the above two discrete transformations, only the $\left\{
\theta\rightarrow-\theta\right\}  $, survives and becomes $\left\{
\Phi\rightarrow-\Phi\right\}  $. Of course Lagrangian (\ref{d.47c}) is
that of the minimally coupled scalar field with constant potential. Finally,
if we assume that $\Psi=\exp\left(  \Phi\right)  $, then the discrete
transformation $\left\{  \Phi\rightarrow-\Phi\right\}  $ becomes a
duality symmetry on the minisuperspace $\left\{  \Psi\rightarrow\Psi
^{-1}\right\}  $. Let us remark that the latter symmetry for
(\ref{d.47c}) and the scale factor duality symmetry for Lagrangian
(\ref{d.40}) are related by the reversal symmetry $\left\{  y\rightarrow
-y\right\}  $ of (\ref{d.33}) in the Cartesian coordinates. 

The admitted common Noether algebra for the Brans-Dicke field
(\ref{d.40}), and the scalar field (\ref{d.47c}) is not surprising
\cite{MTGRG} since the two Lagrangians are related under a conformal
transformation which relates the Jordan and the Einstein frames \cite{cqg}.

\subsection{Duality transformation in $f\left(  R\right)  $-gravity}

In the above dilaton field Lagrangian, we studied the scalar-factor duality.
However, in order to search for discrete transformations which relates
different models, let us consider a spatially flat FLRW spacetime with line
element%
\begin{equation}
ds^{2}=-\frac{1}{N^{2}\left(  t\right)  }dt^{2}+a^{2}\left(  t\right)  \left(
dx^{2}+dy^{2}+dy^{2}\right)  \text{.} \label{d.47}%
\end{equation}
Starting from the O'Hanlon theory \cite{Hanlon}, the equivalent $f\left(
R\right)  $-gravity, in the metric formalism can be easily achieved
\cite{Buda}. Adopting a Lagrange multiplier as in \cite{lan1}, the above
point-like Lagrangian can be suitably recast as
\begin{equation}
L\left(  N,a,\dot{a},\dot{\phi},\dot{\phi}\right)  =N\left(  a,\phi\right)
\left(  6a\phi\dot{a}^{2}+6a\dot{a}\dot{\phi}\right)  +\frac{a^{3}V\left(
\phi\right)  }{N\left(  a,\phi\right)  }, \label{d.48}%
\end{equation}
where we have considered $N\left(  t\right)  =N\left(  a\left(  t\right)
,\phi\left(  t\right)  \right)  $. Furthermore, the relation between the
O'Hanlon scalar field and $f\left(  R\right)  $ gravity is%
\begin{equation}
\phi=f^{\prime}\left(  R\right)  ~~,~~V\left(  \phi\right)  =f^{\prime}\left(
R\right)  R-f, \label{d.49}%
\end{equation}
where the latter first order differential equation is a Clairaut differential
equation which, for specific potentials $V\left(  \phi\right)  $, provides the
functional form of $f\left(  R\right)  $.

Let us assume now that $N\left(  a,\phi\right)  =a^{2\lambda-1}\phi^{\lambda}%
$. The minisuperspace of (\ref{d.48}) admits the maximal dimensional
homothetic algebra. Hence, for potential of the form
\begin{equation}
V\left(  \phi\right)  =V_{0}\phi^{\lambda}\left(  \phi^{\lambda+1}+1\right)
^{-\frac{2\left(  \lambda-2\right)  }{\lambda+1}}~,~\text{for }\lambda\neq-1
\label{d.50}%
\end{equation}
Lagrangian (\ref{d.49}) admits conservation laws which follow from the nonnull
isometry vector
\begin{equation}
X_{f\left(  R\right)  }=a^{-\lambda}\partial_{a}-a^{-1-\lambda}\phi^{-\lambda
}\left(  1+\phi^{1-\lambda}\right)  \partial_{\phi}\,.
\end{equation}
Under the coordinate transformation
\begin{equation}
a=u^{\frac{1}{\lambda+1}}~,~\phi=\left(  uv^{-1}\right)  ^{\frac{1}{\lambda
+1}}\,, \label{d.51}%
\end{equation}
the Lagrangian (\ref{d.48}) becomes%
\begin{equation}
L\left(  u,\dot{u},v,\dot{v}\right)  =\frac{6}{\left(  \lambda+1\right)  ^{2}%
}\dot{u}\dot{v}+V_{0}\left(  u+v\right)  ^{-2\left(  \frac{\lambda-2}%
{\lambda-1}\right)  }\,. \label{d.52}%
\end{equation}
It is not necessary to go in Cartesians coordinates to see that (\ref{d.52})
is invariant under the discrete transformation $\left\{  u\rightarrow
v,~v\rightarrow u\right\}  .~$ Therefore, from (\ref{d.51}), we can see that
the Lagrangian is invariant under the transformation%
\[
a\rightarrow a~,~\phi\rightarrow\phi^{-1}.
\]
We shall say that the solution of the field equations among the two spacetimes%
\begin{equation}
ds^{2}=-\frac{1}{a^{2\lambda-1}\phi^{\lambda}}dt^{2}+a^{2}\left(  t\right)
\left(  dx^{2}+dy^{2}+dy^{2}\right)  \text{,} \label{d.53}%
\end{equation}%
\begin{equation}
ds^{2}=-\frac{\psi^{\lambda}}{a^{2\lambda-1}}dt^{2}+a^{2}\left(  t\right)
\left(  dx^{2}+dy^{2}+dy^{2}\right)  \text{,} \label{d.54}%
\end{equation}
in the O'Hanlon theory, or in $f\left(  R\right)  $ gravity, are related under
the transformation $\phi=\psi^{-1}$, when the potentials of the two theories
$V\left(  \psi\right)  ,~V\left(  \phi\right)  ~$ are related as $V\left(
\phi\right)  =V\left(  \psi^{-1}\right)  $, where $V\left(  \phi\right)  $ is
that of (\ref{d.50}). However, these two models provide the same scale-factor
but in different lapse time. Which means that in the lapse time
${\displaystyle d\tau=\frac{1}{N}dt}$, the two theories provide different
scale factors, i.e. cosmological solutions.

On the other hand, this property is useful for theories where the lapse time
is energy dependent, as in the so-called Gravity's Rainbow \cite{rainbow}. In
other words, lapse time can be considered to be variable in terms of the
dynamical quantities of the model. See, for instance, \cite{rainbow1,rainbow2}.

Furthermore, from (\ref{d.49}) and for $\lambda=2$, we can find a closed-form
for $f\left(  R\right)  $. Hence, $f\left(  R\right)  \simeq R^{2}$, which is
a power law $f\left(  R\right)  $ model which provides a de Sitter universe
\cite{BarOtt}. For $\lambda=0$, from (\ref{d.49}), we have the model%
\begin{equation}
f\left(  R\right)  \simeq -R-f_{1}R^{\frac{4}{3}}\text{,} \label{d.55}%
\end{equation}
where the lapse functions in the two metrics (\ref{d.53}) and (\ref{d.54}) are
$N_{1}\left(  a,\phi\right)  =N_{2}\left(  a,\psi\right)  =a.$ It is
interesting to mention that this model has been found previously from the
application of the Killing tensors in the Lagrangian of the field equations
\cite{AnCQG}, and it has been found that describes the Chevallier, Polarski
and Linder (CPL) parametric dark energy model.

The results presented here are different from those of previous studies on the
relation between duality and Noether symmetries, and, in some sense,
generalize them \cite{dualityCap,dualityCap1,dualityCap2}. Specifically, if we
consider the calculations in \cite{dualityCap} and assume the present
approach, we will see that what is there considered a conserved quantity, is
conserved if and only if the $f\left(  R\right)  $ function is going to
vanish. This fact follows from the complete solution of the Noether symmetry
condition adopted here. On the other hand, using the specific approach for
Noether symmetries adopted in \cite{dualityCap} for Lagrangian (\ref{d.48})
with $N=1$, only the $f\left(  R\right)  \simeq R^{3/2}~$ satisfies the
Noether symmetry condition \cite{fr1,fr2}. The problem can be bypassed
assuming a generic $f(R)$ gravity model and changing the point-like Lagrangian
which is no more that of $f\left(  R\right)  $ gravity derived in the metric
formalism. This can be easily seen by comparing the power-law solution found
for $f\left(  R\right)  =\sqrt{R}$ in \cite{dualityCap}, with the power law
solution which follows from (\ref{d.48}) for the same functional form of
$f\left(  R\right)  $ \cite{AnNEB}. Furthermore, by using the classification
for the Noether symmetries in $f\left(  R\right)  $ gravity for the lapse time
$N=1$, as in \cite{AnNEB}, we can see that the scale factor duality, which
follows from Noether point symmetries, does not exist in $f\left(  R\right)  $
gravity. As a final comment, we have to say that the main role in determining
the symmetries for $f(R)$ gravity is played by the lapse function $N$ and the
form of the point-like Lagrangian adopted\footnote{For a discussion see also
\cite{dimfr}.}. The above results are general and results in \cite{dualityCap}
can be easily framed in this approach being careful with the above remarks.

\section{Conclusions}

\label{conclusion}

The relation between discrete transformations and Noether (point) symmetries
for cosmological models in the context of the minisuperspace has been
discussed. Using results from differential geometry, Noether (point)
symmetries are generated by the elements of the Homothetic algebra of the
given minisuperspace. We discussed cosmological dynamical systems that admit a
Noether symmetry in the normal coordinates of the Noether symmetry vector.
Furthermore, in the normal coordinates, we studied the existence of discrete
transformations. We applied that results and we showed that Brans-Dicke-like
cosmological models with linear potentials admit a discrete transformation
which keeps the field equations invariant, while when the Brans-Dicke
parameter is $\omega_{BD}=1$, the discrete transformation is a scale-factor
duality transformation and the Brans-Dicke scalar field is equivalent with the
dilaton scalar field. Finally, for arbitrary Brans-Dicke parameter, a discrete
transformation which leaves invariant the field equations has been found.

Furthermore we showed that there exist a family of scalar-field models,
equivalent in O'Hanlon gravity and in $f\left(  R\right)  $-gravity, where the
solutions of the models are related under discrete transformations for the
scalar field. A particular transformation is a duality transformation on the minisuperspace.

In particular what we showed is that the existence of a Noether
symmetry for the gravitational Lagrangian can provide discrete symmetries for
the field equations. These discrete symmetries are reversal transformations in
the normal coordinates of the symmetry vector, while in other coordinates,
when they survive, can give another behavior for the original system. 

Indeed the Brans-Dicke field with linear potential, consequently the
dilaton field, and the minimally coupled scalar field with constant potential
are maximally symmetric and describe the same classical mechanical system of
the \textquotedblleft oscillator\textquotedblright\ in the two dimensional
flat space with Lorentzian signature. However, while the reversal symmetry
$\left\{  y\rightarrow-y\right\}  $, for the \textquotedblleft
oscillator \textquotedblright\ , becomes a scale-factor duality symmetry for
the dilaton field, or more general, the discrete symmetry (\ref{d.45}%
)-(\ref{d.46}) for the Brans-Dicke field for the minimally coupled field, the
reversal symmetry is again a reversal symmetry for the scalar field. 

An issue that we did not discussed in details on that work is what happens as
far as concerns the scale-factor duality transformation under conformal
transformations. As we mentioned above in the case of the dilaton field with
action integral (\ref{d.11}) defined in the Jordan frame, in the Einstein
frame is equivalent to that of a minimally coupled scalar field with constant
potentials, where, obviously, the field equations does not admit scale-factor
duality symmetry but the latter becomes a reversal symmetry for the scalar
field. The reason for that is that\ Noether symmetries survive under conformal
transformations \cite{MTGRG,dimfr,cqg}. Hence, in order to explore the
relation between point transformations and discrete transformations in
conformal equivalent theories, in a forthcoming work we will extend this
analysis to the Lie point symmetries of the Wheeler-DeWitt equation.

\begin{acknowledgments}
The authors would like to thank the anonymous referee for his/her careful
reading and for his/her valuable comments which improved the quality and the
presentation of this work.\textbf{ }The research of AP was supported by
FONDECYT postdoctoral grant no. 3160121. SC acknowledges the support of INFN
(\textit{iniziative specifiche} QGSKY and TEONGRAV). This article is based upon work from COST Action (CANTATA/CA15117), supported by COST (European Cooperation in Science and
Technology).
\end{acknowledgments}

\end{document}